# An Analytical Perspective to Traffic Engineering in Anonymous Communication Systems

Mehran Alidoost Nia
Department of Computer Engineering
University of Guilan
Rasht, Iran
alidoost@msc.guilan.ac.ir

Eduard Babulak
Computer Science & Engineering
Maharishi University of Management
Iowa, USA
babulak@ieee.org

Benjamin Fabian
Information Systems
Hochschule für Telekommunikation
Leipzig, Germany
fabian@hft-leipzig.de

Reza Ebrahimi Atani
Department of Computer Engineering
University of Guilan
Rasht, Iran
rebrahimi@ guilan.ac.ir

*Abstract—* **Anonymous communication systems (ACS) offer privacy and anonymity through the Internet. They are mostly free tools and are popular among users all over the world. In the recent years, anonymity applications faced many problems regarding traffic engineering methods. Even though they ensure privacy under some conditions, their anonymity will be endangered by high performance processing units. To address these issues, this study is devoted to investigating traffic-engineering methods in anonymous communication systems, and proposes an analytical view of the current issues in ACS privacy and anonymity. Our study also indicates new types of solutions for these current issues with ACS.**

## I. Introduction

In the recent years, cyber systems faced new challenges regarding user's privacy. Various communication applications have emerged, the number of Internet users is growing rapidly, and connecting facilities are increasing respectively. At the moment, the number of cyber threats is growing very fast. The main problem is caused by inadvertent network users who do not follow security rules for their network-related activities. Users seek for practical solutions to ensure their network-based activities are safe and their private data would not be shared with third parties. One of the straightforward solutions to this demand is to use anonymous communication systems (ACS). They offer free applications but the interference of governmental organizations decreases their credibility [1]. ACS is a good response to this type of issues but there is one basic question. We are sure that they are not fully invulnerable. This paper is an attempt to indicate new challenges in security of ACS.

Anonymous communication systems are used to preserve user's privacy and anonymity. They provide privacy using strong cryptography methods [1]. Multi-layered encryption made ACS stronger than ordinary encryption and by the way, the power of Mix Networking is given to it [2]. So, onion routers and special protocols help to provide layered encryption [3]. Consider a network structure in which its source only knows about the next station. Each station has its own encryption layer. After at least three encrypted steps, it is very hard to detect the location of the first station. If we name stations as onion routers, it will be an ordinary ACS. Tor is the most salient ACS in the world which is very popular among Internet users [4]. It provides free access using volunteer middle servers. This part of ACS' architecture increases anonymity of the entire user base.

The main issue regarding anonymous communication systems is the possibility of performing traffic engineering in such applications. It would be necessary to investigate every direction which may undermine privacy and security of the users. It is very difficult to address all of the vulnerabilities and threats. However, in this paper we aim to indicate the most popular approaches that endanger security and privacy of ACS. An analytical view is given in two different directions. The first is to investigate the possible and current approaches in traffic engineering that may have a destructive effect on ACS performance and security. The second is related to the new solutions and alternatives for current anonymous channels used in popular anonymity applications like Tor. Here, we discuss the pros and cons of the mentioned approaches via an analytical perspective and practical evaluations.

New approaches are proposed when the users feel uncertain about security mechanisms of the system. Security reports and social developments escalate this situation. Security engineers must have alternative solutions and security enhancements to provide users with high immunity and privacy. For example, recent reports show that some governmental activities have started to reveal Tor's network traffic [1]. This would not be possible without using high performance and distributed computing systems. To address this issue, some researchers devoted their research work to design novel anonymous channels [5]. These new channels change the current order of anonymous applications. Some research works have shown hidden defects and security issues in anonymous communication systems [6]. They help researchers to understand current situations and real threats. So, they would design and prepare new solutions to cope with such security problems.

Traffic engineering is a branch in network systems which is devoted to analyzing a sequence of network packets [7]. In traffic engineering, instance data, consisting of a set of packets, is sampled [8]. This data is categorized and compared to pre-known signatures or a knowledge database. If any match is found, that traffic flow is considered as a security exception for a security exception. Otherwise, ordinary transactions are continued. In this process, the main goal is to find some patterns of network traffic which may cause potential threats.

The rest of the paper is organized as follows: current traffic engineering approaches are presented in Section II. New solutions and alternative systems are illustrated in Section III. The analysis, practical results, comparison, and the evaluation are investigated in Section IV. Finally, the paper is concluded in Section V.

## II. TRAFFIC ENGINEERING APPROACHES IN ACS

By definitions, Mix networks adopt network routing protocols which are more complex than ordinary routing and consequently, they are harder to trace via third parties [9]. They mostly use several proxies between the sources and their destinations. Anonymous communication systems are part of applications that evolved from Mix networks.

### A. Strong Cryptography Methods in ACS

Tor is the most popular ACS which provides a high level anonymity and privacy for users throughout the Internet [10]. Typically, ACS supports associated applications with two basic features including encryption of all transmitted data packets and hiding user's identity. Using strong encryption methods, it enhances security and privacy of the user's network identity and ensures users with high level of immunity against common attacks. Originally, a series of volunteer middle servers called relays, helps to provide mix networks through the connections. Each server may encrypt or decrypt its own layer of security and overall, they make it hard to reveal the origin of the senders. This procedure enables the system to provide anonymity feature to its applications.

As mentioned, Tor is the most popular ACS among Internet users. In the recent years, it has been known as the most immune anonymous application for private and secure network interactions. But there are new issues regarding its privacy-preserving features and performance that make users concerned about their privacy [11]. According to many works, Tor causes performance issues when real-time communications matter [12]. This issue is tangible because Tor uses a series of middle-servers to support onion routing protocols [13]. It includes some proxies, encryptions, decryptions, and key exchange protocols. In fact, performance issues of Tor effect its security architecture.

Tor supports full encryption and decryption methods using strong cryptographic algorithms. It is not acceptable if someone would be able to reach plain data within the Tor network. But recent security analysis shows that Tor and similar ACS applications may have several vulnerabilities leading to information leakage or other types of exploits [14]. As shown in Figure 1, its cryptography method is based on at least three different layers between sender and receivers. To send data from Alice to an ACS node such as an exit node (Dan), we need to pass at least the following encryption and relay steps:

*i. Relay CAB AESK1(AESK2(AESK3(Data or commands)))*

*ii. Relay CBC AESK2(AESK3(Data or commands))*

*iii. Relay CCD AESK3(Data or commands)*

*iv. Decrypt and run Data or commands on last OR*

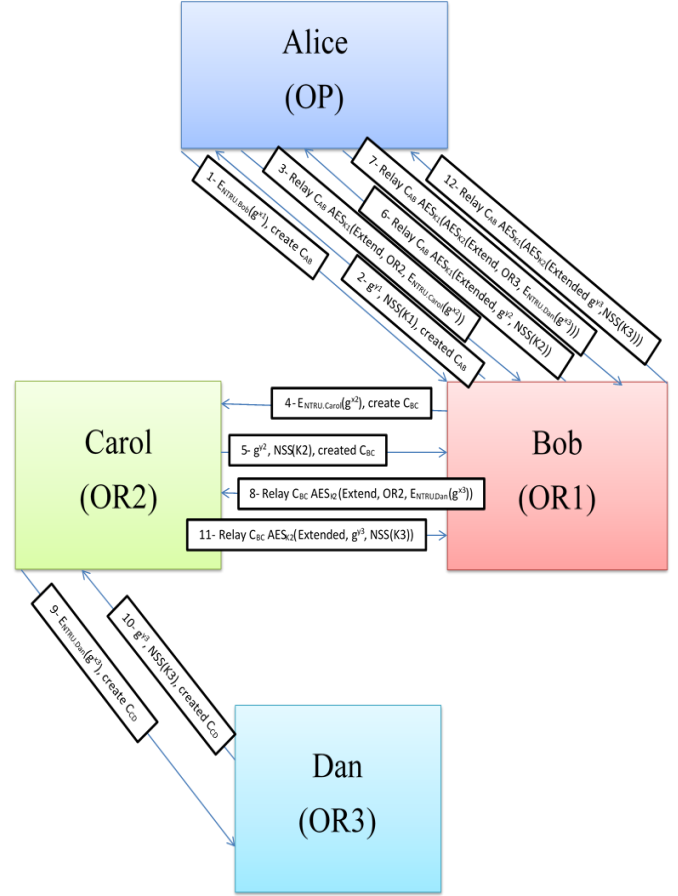

Fig. 1. Layered cryptography methods used in popular ACS security architectures.

During the same steps, plain data is encrypted using three different keys. Each layer decrypts its related layer and cannot not access plain data and its origin. Starting from Dan, the following steps are conducted for anwering in order to complete the process of anonymous communication.

*i. Relay CCD AESK3(results)*

*ii. Relay CBC AESK2(AESK3(results))*

*iii. Relay CAB AESK1(AESK2(AESK3(results)))*

*iv. Decrypt and show results on OP*

Consider that even with the described strong cryptographic methods, there are still some kinds of vulnerabilities which arise from traffic engineering methods. In the following section, we describe related current issues.

## B. Traffic Engineering Methods

Discerning new attacks and common vulnerabilities could help to improve security aspects of ACSs before any exploit happens. A considerable part of the recent research on ACS is devoted to detecting these vulnerabilities. The main tool for detecting an anomaly is traffic engineering. Traffic engineering is used in both detecting a new threat through the network interactions and discerning the behavior of a specific application in the network. The first aspect aids the security of the anonymous communication system. But the second may jeopardize its security infrastructure. A prerequisite for this attempt is the ability to discern encrypted data through the Internet. One of viable approaches is Hidden Markov Model which firstly provided detection of a network flow among an encrypted proxy [15]. Most of the network analysis techniques are based on the three vital origins: direction, time and packet size. Some techniques like the HMM [15] and session classification [16] consider time and size features for their traffic engineering methods, but some others use the direction of the data flow [17] or mixed information of the direction and size or time [18].

Because of their anonymous nature, they should be less dependent on directions [19]. Allegedly, an important source of security bugs related to user's privacy is browser fingerprinting on the web [20]. It leads to executing some analysis attack on anonymous applications [21]. Internet traffic analysis requires super-computing units that are placed in the Internet backbone [22]. This is the main concern in security of these systems.

As long as we can detect the vulnerabilities, we will prevent both active and passive attacks such as Sybil attack, timing, packet counting, profiling, disclosure, forgery, deanonymization, and intersection attacks [23]. Consider that most of these attacks are categorized as public vulnerabilities which are rampant among all types of network applications. Some vulnerabilities are associated to special types of application's architecture and model. For example, when an ACS is prepared to work via a peer-to-peer network, it is obvious that all types of attacks and vulnerabilities associated with the host environment could affect the anonymous application.

One of our recent investigations in traffic engineering has led to detecting unidentified network traffic [24]. In the same work, we found out when a new type of network flow has appeared, we could detect its nature. However, even though we could not fully analyze its danger level because of strong cryptography methods, the ability to estimate the type of application behind the same traffic still exists. This means that even if we could not access plain data of an anonymous communication system like Tor, we can understand its type and that it is Tor traffic. This is the source of many problems in ACS security.

The proposed manner to analyze cryptographic data is as follows. Firstly, we get a set of sampled data from real network traffic. If it is matched to any pre-known samples in the database, it is detected. If not, we can set a new record for the same traffic and guess its danger level according to the other vulnerabilities. When the subject is an anonymous communication system, we can perform comparisons to determine a most precise similarity level except danger level. To do that, we have to use a random process. We chose random walk that uses type theory as its final similarity measurement [25].

To establish the process of random walk, let $R_k$ be a set of random vectors, which are considered to be independent. $R_k$ only takes values $\pm e_i$, $i=1,2,...,d$ with probability of $1/2d$ where $d$ is the maximum distance. The challenge is to monitor the walker's position after $n$ steps, while performing calculation of equation (1) where $X$ represents each step. This is called simple random walk and one of its applications is to simulate random behaviors in the real world, especially in computer networks [26].

$$S_n = \sum_{k=1}^{n} X_k \qquad (1)$$

When random walk is used in modeling of the system, we let a random walker move along, using several modeling parameters. Here, time and size are considered as two basic parameters. This gives similarity measurements for several input network traffics. If one of them is a set of sequences for Tor's real traffic, we could measure similarity of each input sequence with that. The work has a complex method to perform the process of sampling, categorizing, estimating, modeling, and similarity measuring in the entire system.

## III. NEW SOLUTIONS AND ALTERNATIVES

As mentioned before, traffic engineering poses a potential threat on the security of anonymous communication systems. To counter this threat, there are several research works devoted to address this issue. Some of them tried to cover security defects by either improving or restricting a specific part of ACS. But some of them try to design anonymous channels from scratch. One of the appropriate responses to this issue is behind current research works on developing a new generation of ACSs [27]. It is very important to consider new designs but also the improvement of classic ACSs in order to reduce current vulnerabilities.

The second solution is better because traffic engineering aims to work through anonymous channels. So, to address this issue, we must focus to change the behavior of the network-related part of the system. Changing the way how an ACS produces network traffic is useful. Also, it is best to construct a network traffic pattern which is very similar to other ordinary applications like Skype and Bittorent. On the other hand, it causes intruders to make a mistake in discerning anonymous communications system's network traffic.

Making changes in network traffic pattern is very complex. It would be regulated by both senders and receivers. In one of our research works, we have designed a new anonymous channel which is perfect to counter traffic engineering methods [5]. We have sampled some ordinary network traffic which belongs to specific applications and protocols like Bittorent, Skype, POP3, HTTP, SSH, SMTP and etc. A regulator for both sides is designed using crypto types [28]. Crypto types are responsible for producing a network pattern with time and size related to a specific application. These crypto types are shared between sender and receiver and the encrypted flow will be reconstructed on the receiver's side. In each step, random walk helps to choose a randomly selected direction which represents any type of sampled applications.

The result of that work shows that when an intruder tries to use traffic engineering tricks to discern the type of network flow, the sampled sequence of the packets indicates that it does belong to an ordinary applications like Skype. So, the intruder will give up and follow another sequence. We will be surprised when the second network traffic which is produced in the next session flows through the network. If the intruder sampled a set of packets from the previous flow, the new one will not be matched with the previous network traffic. It is because in each step, random walk changes the traffic pattern. If the size of produced pattern is very high, the probability of detecting a similar pattern in even 1000 times repetition of sampling is inconsiderable.

$$P_{s=0}(w) = \prod_{i=1}^{m}[\frac{1}{(2 \times d) - l}] \quad (2)$$

It is possible to calculate the probability of discerning the entire walk which is constructed of $m$ points that should be correctly matched. Equation 2 indicates this, where $m$ is the number of walks that is processed and $d$ is the same dimension. In self-avoiding random walk, probability is measured in a different way. This means that some directions should be easily removed from the set of adjacent nodes. The equation shows the probability of determining match points with self-avoiding random walk where $l$ is the set of visited nodes. The intruder does not understand the policies which are executed and this constraint should be considered as a game changer parameter. We called this technique *Multimodality Injection*. In equation 2, $s$ represents the start point of the walker which is set zero by default.

The other type of solutions is about designing a new ACS. Even though Tor is very popular, it is not the only origin of ACS researches. Mixnet is one of the ACSs which focuses on traffic analysis and recently is used to develop new research to prevent selective DoS attacks [29]. Also, it has the ability to incorporate with Lattice based cryptography system to ensure privacy of location in mobile systems [31]. As the other ACS, we can refer to ALERT [11], Anonymizer [11], Crowds [11], Herd [27], and PACOM [11].

IV. ANALYSIS, COMPARISON, AND EVALUATION

As mentioned in the earlier sections, we showed pros and cons of anonymous communications systems according to current trends especially those related to traffic engineering approaches. In this section, we aim to analyze the mentioned threats and solutions to make it clear whether an ACS could be fully secured against traffic engineering attacks or not.

*A. Security Analysis and Comparison*

The fact that anonymity applications are susceptible to traffic-engineering attacks is clear enough to consider it as a potential threat. Based on the performed method, the danger level may be variant and depends on sampled data. We divide the possible situations into two categories. The first includes some traffic engineering methods that aim to detect encrypted network flows. The second is about detecting behavior of the application without considering its encrypted contents. As we mentioned earlier, an example for the first approach is Hidden Markov Model [15]. The second manner is very general and seems independent from anonymity-specific attributes like hidden location and encrypted contents. This type of traffic engineering methods would be used in all network flow samples. They mostly depend on the behavior of applications. Two of the recent approaches are session level flow classification [16] and detection of unidentified network flows [24]. As shown in Table 1, some of recent approaches in traffic engineering are compared using variety of parameters.

Table 1- Categorization of traffic engineering techniques by types and features.

| Technique | Average True Detection Rate | Type of Traffic Engineering Technique | Complexity | Ability to work Online |
|---|---|---|---|---|
| Unidentified Network Flow Detection [24] | 95% | behavioral | $x^4$ | Yes |
| Session Level [16] | 99% | behavioral | $e^x$ | Yes |
| Co-clustering [30] | 93% | behavioral | $e^x$ | Yes |
| HMM [15] | 80% | content-based | $e^x$ | Yes |
| EPM [32] | 90% | content-based | $nx$ | No |
| Distance based [30] | 80% | behavioral | $x^4$ | Yes |

According to the Table 1, traffic-engineering methods that can be used to identify anonymous communication systems are categorized by the type of techniques. The accuracy of detection in behavioral methods is naturally higher than content-based methods. Implementations of behavioral systems have higher complexity and usually perform in cooperation with online monitoring tools. But content-based approaches are good to identify types of cryptographic methods used in that ACS. The traffic sampling in the both of them is identical except for situations when we need to analyze sampled data in offline mode. Typically, content-based methods are appropriate for analyzing the behavior of subsystems like security methods, type of transmitted data, communication protocols, and anomaly detection.

Behavioral traffic-engineering methods map current network level attributes to logical and behavioral parameters. It means that it is not important for them to know what type of contents are transmitted via the network. They only make use of three significant parameters including time, size, and direction. As shown in Figure 2, a visual comparison of mentioned approaches in traffic engineering of ACS is presented. It is obvious that HMM, as one of the best content-based methods, is less accurate than behavioral methods in true detection of an ACS application. So, when it comes to anonymity applications, it is better to consider behavioral traffic engineering methods as potential threats since the use of strong cryptography methods makes an ACS impenetrable against content-based attacks. In the picture, each node represents level of accuracy. At least we can state that the possibility of such a successful analytical attack against ACS is expected to be low.

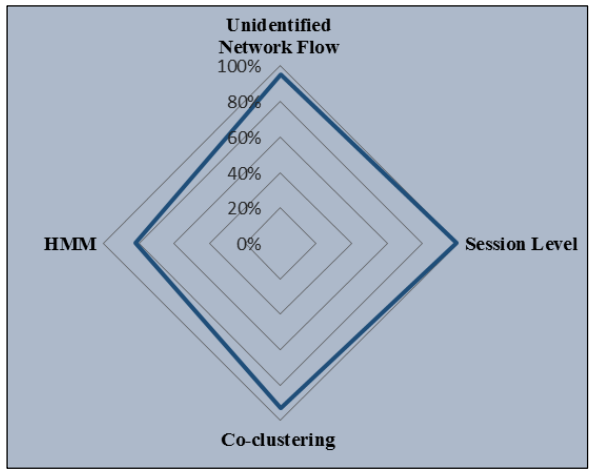

Fig. 2. A visual comparison of the mentioned methods.

*B. Evaluation of Solution*

As mentioned, our investigations show that the main target for any defense strategy in anonymous communication systems should be the behavioral methods. They pose a potential threat to anonymity applications. So, the possible solution is a new structure that makes the ACS strong against behavioral attacks. Our recent research presents a solution to this issue [5]. In that work, a new anonymous channel based on randomization features is offered. It constructs a custom network pattern using random walk. It is very useful to overcome threats reflected by behavioral traffic analysis approaches.

In each generation of RAPS (Random-walk based Anonymous Pattern Simulator), the system produces a new type of pattern based on its pre-organized database. The database is a composition of the most popular applications' sampled data. Based on random steps, the simulator injects a random application's feature to the same generated network flow. When the intruder gets its own sample, in each time, he faces an ordinary type of behaviors. The behavior of applications and protocols like Skype, Bittorent, POP3, HTTP, SSH, Emule etc. is used in the same work to unleash a wrong signal to the intruders.

As long as the variety of sampled applications is enhanced, the immunity against behavioral analysis of traffics is decreased respectively. This work is not the only research dedicated to traffic engineering methods in ACS, but it is one of the creative approaches which concerns analytical attacks. It gives a good view on the ACS issues against traffic engineering methods.

*C. Limitations of Proposal*

This work is conducted using analysis of the current issues in anonymous communication systems according to traffic engineering methods. We have presented some solutions to overcome the mentioned issues. But the solutions are not limited to those stated in this study. We extracted the result of traffic engineering methods from various research works and they have been experimented on using different samples. So, real implementation of these analytical methods may help to receive more precise results. Also, execution on a unique data set definitely gives competitive results.

## V. CONCLUSION AND FUTURE WORK

In this work, we aimed to present a realistic view on current issues of anonymous communication systems where traffic analysis matters. We investigated current challenges and traffic engineering methods that endanger the anonymity and privacy of such applications. We divided traffic analysis methods into two categories including content-based and behavioral. Technical features of these methods were stated using quantitative parameters. The study shows that the main concern to improve ACS against traffic analysis attacks should be focused on behavioral methods. This means that we must propose a solution to change the behavior of an anonymity application via the Internet. A solution is studied to address a new defense strategy against traffic engineering attacks. Our investigations indicate that when it comes to anonymous communication systems, the behavioral methods are stronger than content-based approaches.

Many researchers work on finding and improving vulnerabilities of popular applications such as Tor. But at the same time, new anonymity tools and public applications are emerging which we mentioned in this study. One of the future works is to start a study to analyze other anonymity applications against traffic engineering methods. Another future work is to propose solutions against emerging behavioral methods that aim to reveal ACS applications through the Internet. Each attack needs its corresponding counterattack reaction. Experimenting with all behavioral traffic-engineering methods on the same data set, is considered as another future research work. This would make the power of each method comparable in the same setting.

## ACKNOWLEDGMENT

We would like to appreciate Antonio Ruiz-Martinez for his considerable help and contributions in our previous research works related to anonymous communication systems.